# Exploring the feasibility of Fe(Se,Te) conductors by *ex-situ* Powder-in-Tube method

M. Palombo[1,a)], A. Malagoli[2,b)], M. Pani[1,2], C. Bernini[2], P. Manfrinetti[1,2] A. Palenzona[1,2] and M. Putti[2,3]

[1] *Department of Chemistry and Industrial Chemistry, University of Genova, Via Dodecaneso 31, 16146*

*Genova, Italy*

[2] *CNR-SPIN, Corso Perrone 24, 16152 Genova, Italy*

[3] *Department of Physics, University of Genova, Via Dodecaneso 33, 16146 Genova, Italy*

In this work, the feasibility condition of Powder-In-Tube (PIT) processed wires of Fe(Se,Te) superconductor has been investigated. We faced several technical issues that are extensively described and discussed. In particular, we tested different metals and alloys as external sheaths (Cu, Ag, Nb, Ta, Ni, Fe, cupronickel, brass) concluding that the only sheath that does not affect substantially the Fe(Se,Te) phase is Fe. On the other hand, Fe sheath introduces excess iron in the Fe(Se,Te) phase, which affects the superconducting properties; we investigated the effects of the thermal treatments and of the powder composition in order to avoid it. The maximum *Jc* value obtained in our samples is $4 \cdot 10^2$ A/cm$^2$, comparable to other published values of PIT conductors of the 11 family. We conclude that the fabrication of Fe(Se,Te) wires by PIT method is quite challenging and other approaches should be developed.

**I. INTRODUCTION**

Since the discovery of high temperature superconductivity in the Fe-based superconductors in 2008,[1] huge efforts have been done in several directions going from the research of new superconducting families, to understanding the mechanisms underlying superconductivity and exploring their potential for applications. Fe-based superconductors show extremely large upper critical fields and relatively low electronic anisotropy, which are crucial aspects for power applications.[2,3] However, early polycrystalline materials exhibited strong electromagnetic granularity similar to those observed in cuprate superconductors mainly due to the presence of spurious phases and cracks at the grain boundaries. Despite these problems, in few years great progress has been made improving the phase purity and densification, and the powder-in-tube (PIT) processing method has been implemented for the realization of wires and tapes.[4] This method consists in packing stoichiometric amount of precursor powders into a metal tube, which is subsequently deformed into a wire and heat treated. The PIT technique is extensively used for producing wires and tapes of technical superconductors like Nb$_3$Sn, MgB$_2$ and Bi

---

a) Now at IIS Progress s.r.l., Lungo Bisagno Istria 15, 16141 Genova, Italy

b) A. Malagoli: author to whom correspondence should be addressed. Electronic mail: andrea.malagoli@spin.cnr.it

based cuprates, because it overcomes the problem of mechanical hardness and brittleness of the compounds; moreover, the advantages of using standard equipment and low cost materials make it suitable for industrial development.

PIT wires and tapes of the main three types of Fe-based compounds have been developed. Most of the work has been carried out on the K doped Ba(Sr)Fe$_2$As$_2$ compounds (122 family). This family has fairly large critical temperature, $T_c$, up to 38 K, close to that of MgB$_2$; it is weakly anisotropic, and the superconducting transition does not suffer from significant broadening by thermal fluctuations. Moreover, it is well compatible with silver as sheath and requires low annealing temperature. Indeed, the best transport critical current density, $J_c$, up to $10^5$-$10^6$ A/cm$^2$ in self field, has been obtained so far with the 122 family.[5,6] Recently, $J_c$ values above the practical level ($10^5$ A/cm$^2$ at 10 T) have been obtained[7]

F doped ReFeAsO compounds (1111 family) which have with the largest $T_c$, up to 58 K[8] are more problematic. Indeed, they suffer the loss of fluorine as a consequence of both the thermal treatment[9] and the strong interdiffusion with the metal sheaths.[10] Recently, a transport critical current density above $10^4$ A/cm$^2$ at 4.2 K and in self-field has been achieved in 1111 tapes, but it is strongly suppressed by the field.[11]

Despite their lower $T_c$ (about 15 K in bulk materials), iron-chalcogenide Fe(Te,Se) compounds (11 family) are interesting for applications; indeed, the 11 compounds exhibit huge upper critical field (above 50 T)[12] and high critical current ($10^5 \div 10^6$ A/cm$^2$ in self-field), weakly dependent on magnetic field.[13,14] Moreover, the 11 family has the simplest crystal structure, is easy to synthesize and to handle, is stable in air and does not contain toxic elements. These aspects make the 11 compounds very appealing in order to develop an industrial fabrication process of long cables. However, due to the strong reactivity of Fe(Te,Se) phase with metallic sheaths, only few successes in the realization of wires and tapes have been reported up to now.[15, 16, 17, 18, 19, 20] In particular, a diffusion process of the precursors within a Fe sheath has yielded a significant improvement in the transport $J_c$, allowing to obtain $J_c$ values up to $10^3$ A/cm$^2$ in FeSe wires with a $T_c$ = 10K.

On the other hand, we showed that homogeneous and dense Fe(Te,Se) polycrystalline samples with optimal $T_c$=16 K and well interconnected grains, resulting in a $J_c$ of about $10^3$ A/cm$^2$ and nearly independent of the field, can be obtained by combined melting and annealing processes of the FeSe$_{0.5}$Te$_{0.5}$ phase.[21]

An important fact is that an Fe excess easily occurs in samples prepared by the standard synthesis technique.[22] Fe provides electron doping and it is strongly magnetic, which in turn originates local moments that interact with the adjacent Fe layers,[23] acting as a pair breaker.[24] Specifically, crystals with higher Te content usually contain more Fe excess, whose magnitude decreases with the increase in Se doping.[25] In order to avoid the excess Fe several strategies have been devised, going from various post-annealing treatments, in vacuum,[26] in air,[27] in alcoholic beverages,[28] in dilute acid[29] or in O$_2$,[30] to electrochemical reaction techniques,[31,32] to the lowering of Fe in the starting composition.[33] The excess of Fe may be responsible for the unexpected lagged application research, despite the simpler structure, less toxic nature, and easier synthesis technology of the 11 compounds compared with the other iron-based superconductors.

In this paper we probe how to transfer the melting process that succeeded in improving the superconducting properties of the 11 bulk materials to Fe(Se,Te) wires by using the PIT method. With the aim to establish the reliability condition we explore the interaction between the Fe(Se,Te) powders and several metal sheaths, the role of both thermal treatment and phase composition, and finally we clarify the main problems related to this approach.

## II. EXPERIMENTAL

Fe(Se,Te) powders were prepared starting from Fe powder (spherical <10μm, 99.9 wt %, Alfa Aesar), Se powder (99.999 wt %, Alfa Aesar) and Te pieces (99.9999 wt %, Alfa Aesar). The elements were mixed in stoichiometric ratio, grinded together in a mortar and inserted in a Pyrex™ glass phial. The phial was then sealed under vacuum and placed in a resistance furnace for the synthesis heat-treatment at 500°C for 50 hours. The so-obtained sample was analysed by XRD diffraction with a powder diffractometer (Bragg-Brentano geometry and Ni-filtered Cu K$\alpha$ radiation, $\lambda$=1.54178 Å). The Pearson's Crystal Database was used as library of known inorganic substances.[34]

Selected samples were subjected to thermal analysis, in order to determine the liquidus temperature and, if possible, the formation type of the phase. About 0.8 g of the starting powders were pressed into pellets and closed by arc-welding inside a Mo or Fe crucible. Heating and cooling cycles were done at the rate of 5-10 °C/min by means of a NETZSCH 404S DTA equipment; the temperature was measured to an accuracy of ± 5°C.

Superconducting wires were realised by PIT method, in which a metallic tube was filled with Fe(Se,Te) powders and cold worked through drawing and groove-rolling process to obtain round or square wires of about 1 mm in size. Finally, 8 cm long wire samples underwent a sintering heat treatment in a three-zone tubular furnace with a homogeneity zone (± 0.5 °C) of 16 cm. The maximum temperature was in the range of 600 – 850 °C for about 1 hour. Several metallic sheaths were used and tested with the aim to establish which metal or alloy was more suitable in terms of mechanical and chemical properties for this kind of superconductor. In particular, we tested tubes of Cu, Ag, Nb, Ta, Ni, Fe, cupronickel 90-10 and brass.

The resistivity measurements on the wires were performed with a DC-four-probe system with variable temperature from 300 to 4.2 K. Critical current $I_C$ measurements were performed by a DC-four-probe home-made system in liquid helium bath (4.2 K) and self-field. The micrographic characterizations and analyses were carried out by a SEM CAMBRIDGE S360, coupled with an Energy Dispersive Spectrometer for chemical analysis, EDS: OXFORD Link Pentafet. The acquired images were analysed by ImageJ™ software.[35] In particular, the area fraction of the different phases relatively to the whole cross section of the samples was evaluated. In order to obtain reasonable measurements, different analyses were performed on SEM-BSE images at different magnifications covering the whole area of the Fe(Se,Te) compound inside the wire; in this way it was possible to get an error as low as 5%.

## III. RESULTS
### A. The choice of the sheath

The first step was the selection of suitable materials for the sheath of the superconducting Fe(Se,Te) wires. In the literature an exhaustive study on metals suitable for such a purpose is not present yet, in particular on their behavior in terms of chemical compatibility with the superconducting Fe(Se, Te) phase. The published works report only the use of Fe as metal sheath. Other metals non ferromagnetic and with higher thermal conductivity are preferable to Fe, but no information about the use of other sheaths is available. In the selection of a suitable sheath materials, the following characteristics are also crucial: melting temperature higher than the required heat treatment temperature (in this work the heat treatments were performed at a maximum temperature of 850 °C), ductility, and chemical compatibility with the superconducting phase. Therefore, we tested several metals and alloys, easily commercially available, either coupled or singularly: Cu, Ag, Nb, Ta, Ni, cupronickel 90-10 and brass. The effects induced by the sheath on the superconducting phase were investigated after the heat treatment carried out at temperatures ranging from 700 and 850 °C, which are typical temperatures used for the sintering of the Fe(Se, Te) powders. The results obtained are described in the following.

Figure 1 shows details of the cross-section of some representative wires made with different metal sheaths, filled with the same $FeSe_{0.5}Te_{0.5}$ powder.

The silver sheath has been successfully used as metal sheath in Bi-based cuprates wires and in the 122 iron-based superconductors. Thus, as a first attempt, Ag barriers in Fe-sheathed (figure 1a) and Cu sheathed (figure 1b) wires has been tested. It comes out that Ag has a strong influence on the Fe(Se,Te) compound; in fact, after the heat treatment, both Ag and Cu (Figures 1a and 1b, respectively) diffuse up to the middle of the wire generating ternary and quaternary compounds with Fe, Se and Te and degrading large amounts of powder.

Fig. 1c shows a Ni-sheathed tape, heat-treated at 850°C: it can be observed that nickel subtracts iron from the Fe(Se,Te) phase, producing a solid solution with composition of about Fe 98 at.% and Ni 2 at.%, as revealed by EDS analysis. The Fe-Ni spots are easily recognizable by their squared-shape. Nickel pollution already starts with heat treatments carried out at 700°C.

Fig. 1d shows a Nb-sheathed wire, heat-treated at 850°C; niobium develops two different interface reaction layers between sheath and powder; the nearest to the sheath is basically a Fe-rich niobium alloy, whereas the second is a Se-rich layer which includes, in addition to niobium, both tellurium and iron. Moreover, Nb enters into the main phase, which in turn is enriched in tellurium. In addition to the unbalancing of the Fe(Se,Te) compound, some random and large Fe-islands with the size of tens of microns are observed. It is interesting to mention that Nb/FeSe and Nb/FeTe interfaces have been investigated before and after the heat treatment at 584 °C.[36][37] No reaction layers have been observed after the heat treatment performed at such temperature which is indeed too low to produce connectivity between the grains.

Fig. 1e shows a cupronickel 90-10-sheathed wire with tantalum barrier, heat-treated at 700°C: tantalum creates an interface layer between sheath and powder polluting it; its composition includes all the elements of the Fe(Se,Te) phase and also other elements, if present (e.g. double sheathed wires, with broken Ta barrier). Cupronickel has a very degrading effect on the powder; as for nickel, its damaging effects are visible already after heat treatments carried out at 700°C.

Fig. 1f shows a brass-sheathed wire, heat-treated at 850°C: the used brass (ISO CuZn35) is the most commercially diffused, an alloy of copper (65 wt. %) and (zinc 35 wt. %). This copper-alloy, similarly to niobium, originates two different layers. The outer one is composed of a quaternary Fe-Zn-Se-Te phase rich in Zn, while the inner one is made of a Cu-rich phase; well visible islands with the same composition of the latter are immersed in the main phase. The original Fe(Se,Te) powder appears as completely copper-polluted.

In conclusion, we found that, although the tested metal sheaths are largely used in superconducting wire production, almost all of them react with the Fe(Se,Te) compound inside the wire, damaging it.

Finally, Fe-sheathed wires have been prepared. Fig. 2 shows the perspective view of the cold-worked wire by groove rolling and drawing down to a round section with diameter of 2.55 mm; after machining, the wire was heated to 850°C in Ar atmosphere. The corresponding micrographic cross section is reported at different magnifications.

In this case, since Fe is a constituting element, the Fe(Se,Te) phase is not destroyed. However, as evidenced by the different grey scale intensity of the back-scattered images, a de-mixing in two compositions occurs. These compositions correspond to a dark Se-rich phase and to a bright Te-rich phase, respectively labelled "D" and "B" in Fig. 2. Similar de-mixing has been observed in polycrystalline $FeSe_{0.5}Te_{0.5}$ samples after melting,[21,33] and it has been partially recovered by applying suitable heat treatments.

The conclusion of this section is that iron is the only metal sheath compatible with the Fe(Se,Te) phase. From a chemical point of view, unlike the other analyzed metals, it does not react with the superconducting phase to give different new phases; on the contrary, the tetragonal Fe(Se,Te) phase is maintained, although with compositional modifications and phase de-mixing. The role of the heat treatments, also in view of phase homogenization, is discussed in the next section.

**B. Heat treatments**

In order to investigate how to favour the grain connectivity, avoiding the phase de-mixing, if possible, different heat treatments (HTs) have been performed.

From DTA analyses the melting temperatures of the $FeSe_{0.5}Te_{0.5}$ and $Fe_{0.95}Se_{0.65}Te_{0.35}$ compounds were established to be ~ 835 and 810°C, respectively; thus, different thermal treatments were planned at selected temperatures above and below the corresponding melting point.

In figure 3 SEM images of Fe-sheathed wires containing FeSe$_{0.5}$Te$_{0.5}$ powder treated at 700°, 800° and 850° C, are reported.

The 700°C-HT wire presents the typical aspect of a sintered material with separated grains of average size 10-20 µm; some iron oxides and little squared Fe-spots are visible too. The same microstructure can be noted in wires treated at 800°C, where again some coarsened and randomly disposed Fe-spots are clearly visible. In these two wires the presence of the dark phase is hard to distinguish because of its inter-granular arrangement; furthermore, because of the light coalescence, it can be mistaken with Fe spots or oxides.

The 850°C-HT wire exhibits a different microstructure, which proves that at this temperature the melting of the phase has occurred. Both the dark and the bright phases are clearly noticeable: the main bright phase forms primary grains with rounded shape surrounded by the dark phase. In this case, the grain dimension has been estimated around 50÷100 µm (Fig. 3).

The EDS analyses were carried out on the whole section of the Fe(Se,Te) filaments subjected to different HTs. For each sample, the results showed that: (i) both the bright and the dark phases are always present, and are characterised by different chemical compositions. In particular, the bright phase (majority phase) shows a ratio [Se]/[Te] < 1, together with a Fe content higher than in the starting powder, being Fe$_{1.03(1)}$Se$_{0.44(4)}$Te$_{0.56(4)}$ its typical composition; the dark phase (minority phase) is Se-rich and its composition, close to FeSe$_{0.9}$Te$_{0.1}$, varies very little. (ii) The global average composition differs from that of the starting powder, being unbalanced at the expense of Se and with an excess of Fe (Fe$_{1.04(2)}$Se$_{0.48(6)}$Te$_{0.52(6)}$).

The unbalancing is independent of the HT; the loss of Se can be due to the iron sheath permeability to this element, the excess of iron can come from the sheath itself.

**C. Composition of the Starting Powder**

In order to avoid the phase unbalancing characterized by the Se deficiency and Fe excess, powders with different starting compositions were prepared. It is worth noting that the *FeSe$_x$Te$_{1-x}$* phase diagram with x> 0.5 is not well known, indeed, polycrystalline samples with Se concentrations in the range of 0.6<x<0.8 are found to be multi-phase.[38,39] On the other hand, increasing the Se content may have two advantages which counterbalance the negative effects due to the extra Fe coming from the sheath: (i) the Fe excess, naturally included in the tetragonal phase, decreases [25]; (ii) the extreme composition (x=1) accepts excess iron, being superconducting only in the composition FeSe$_{0.82}$.[40] Thus, we prepared four different powder batches in which Fe varied in the range 50-47 at. % and Se in the range 25-34 at. % (see Table 1).

Each batch of powders was characterized by XRD measurements in order to confirm the nominal stoichiometry and to check the homogeneity of the sample. Figure 4 shows the XRD patterns of the as-prepared powders, in comparison with that calculated for FeSe$_{0.5}$Te$_{0.5}$, chosen as reference compound. The results showed that, while samples with 50 at. % Fe are substantially homogeneous, with an average composition in agreement with the nominal one, the samples prepared with a lower content of Fe and a ratio [Se]/[Te] > 1 contain other species along with the main Fe(Se,Te) phase. Indeed, in both SE31 and SE34 powders, the phase Fe(Te$_{1-x}$Se$_x$)$_2$ appears as secondary phase, while sample SE33 contains additional peaks of the hexagonal NiAs-type phase. Taking into account the conditions adopted in the present work for the synthesis, the presence as impurities of these ternary compounds is compatible with the stability regions of the corresponding binary compounds, as reported in the Fe-Se and Fe-Te phase diagrams.[41]

X-ray analyses were also performed on powders extracted from the wires (breaking the iron sheath), after their manufacturing and the subsequent heat treatments. A feature common to all samples with high Se content (SE31, SE33, SE34), is the disappearance of the secondary phases after the heat treatments. In particular, for the Fe(Te$_{1-x}$Se$_x$)$_2$, this can be explained considering that both the (isostructural) compounds FeSe$_2$ and FeTe$_2$ decompose at temperatures higher than 585 and 519° C, respectively. Furthermore, as already described for SE25, also for the samples enriched in selenium, the peaks of the Fe$_{1-y}$Se$_{1-x}$Te$_x$ phase split in two distinct series after the heat treatments, as shown in Figure 5a.

The refinement of the XRD patterns (Figure 5b) shows the coexistence of two phases: one Se-rich with composition close to FeSe$_{0.9}$Te$_{0.1}$ and the other with ratio [Se]/[Te] < 1 (SE25) or [Se]/[Te] > 1  (SE31, SE33, SE34). In the latter case, this ratio varies between 1.1 and 1.4. Similar values are obtained from the corresponding EDS analyses, confirming that the dark phase has the composition Fe$_1$Se$_{0.9}$Te$_{0.1}$ while the bright phase composition varies around Fe$_{1.02(2)}$Se$_{0.56(3)}$Te$_{0.44(3)}$.

The SEM-BSE images, limited to the two highest temperatures (800 and 850 ° C), are collected in Figure 6. Considering the microstructure, it can be noted that the dark phase coalescence grows as the HT temperature is increased. Moreover, in the melted wires (850° C), the two phases appear clearly separated, and lamellar eutectics, together with iron-rich precipitates, are visible near the boundaries of the dark phase. By increasing the Se amount, larger regions of the dark phase are observed; they are characterized by rounded, flower-shaped boundaries, suggesting that the dark phase melts at a lower temperature than the bright one. Indeed these regions appear dendrite-like shaped and mainly oriented along the longitudinal axis of the wire.

The acquired images were analysed in order to evaluate the relative percentage of the two phases: while the ratio between dark and bright phase remains constant (for a given sample) as the HT temperature increases, the dark phase percentage increases with the increase in the starting Se content. The results, reported in Table 2, indicate that the dark phase percentage varies from 15÷20% for the SE25 wires to 30÷35% for the SE34 wires. Table 2 also reports the global chemical composition evaluated by EDS on the whole cross-section of the wire filament for each starting powder, averaging on wire treated at 800° and 850°C (samples of fig.6). It can be noted that the Fe content is around of 52-54%, larger than that of the starting powders, independently of the initial composition.

TABLE 1. Atomic percentage content for iron, selenium and tellurium related to the empirical formula and to the label of each prepared powder.

| *Powder label* | *Empirical formula* | *Atomic content [%]* | | |
|---|---|---|---|---|
| | | *Fe* | *Se* | *Te* |
| *SE25* | $FeSe_{0.5}Te_{0.5}$ | *50.0* | *25.0* | *25.0* |
| *SE31* | $Fe_{0.95}Se_{0.60}Te_{0.40}$ | *48.7* | *30.8* | *20.5* |
| *SE33* | $Fe_{0.95}Se_{0.65}Te_{0.35}$ | *48.7* | *33.3* | *18.0* |
| *SE34* | $Fe_{0.90}Se_{0.65}Te_{0.35}$ | *47.4* | *34.2* | *18.4* |

TABLE 2. Dark phase percentage and global chemical composition evaluated by EDS for the wires shown in Fig. 6. These data are obtained by averaging on the filament cross-section.

| *Wires* | *Global composition [at. %]* | | | *Dark phase percentage* |
|---|---|---|---|---|
| | *Fe ($\sigma$)* | *Se ($\sigma$)* | *Te ($\sigma$)* | |
| *SE25 wires* | *53 (0.5)* | *23 (0.5)* | *24 (0.5)* | *15÷20 %* |
| *SE31 wires* | *53 (0.7)* | *28 (1.2)* | *19 (1.2)* | *20÷25 %* |
| *SE33 wires* | *54 (1.4)* | *29 (0.7)* | *17 (0.7)* | *25÷30 %* |
| *SE34 wires* | *52 (0.6)* | *31 (0.1)* | *17 (0.6)* | *30÷35 %* |

The resistive transitions of the considered wires are also reported in Figure 6: all the curves show nearly the same $T_c$ onset at about 8-9 K, regardless of starting powders and HT. However, a complete transition is observed only for SE31, SE33 and SE34 wires, both sintered and melted. The best results are obtained for the SE33 melted wire, for which the $T_c$ onset is 9.3 K and zero resistivity is reached at 8 K.

The critical current has been evaluated in all the melted wires (850° C), and the V-I curves are reported in Figure 7: from these curves we evaluated $I_c$ = 30, 120, 240 and 200 mA, for the SE25, SE30, SE31 and SE34 wires, respectively. The critical current density values $J_{c\_TOT}$, estimated by considering the total cross-section of the Fe(Te,Se) filament, are in the range 12-120 A/cm$^2$ (see table 3). These values are quite low, and tend to increase with increasing Se content.

TABLE 3. Critical current densities evaluated for the curves of the wires melted at 850°C shown in Fig. 7; $J_{c\_TOT}$ indicates the critical current density evaluated by considering the whole wire cross-section excluding the sheath, whereas $J_{c\_D}$ was evaluated just considering the dark phase percentage values shown in Table 2.

| Melted wire | $I_c$ (mA) | $J_{c\_TOT}$ (A/cm$^2$) | $J_{c\_D}$ (A/cm$^2$) |
|---|---|---|---|
| SE25 | 30 | 12 | 60-80 |
| SE31 | 120 | 50 | 200-250 |
| SE33 | 240 | 100 | 333-400 |
| SE34 | 200 | 80 | 230-270 |

## IV. DISCUSSION

The aim of this work is to explore the feasibility of Fe(Se,Te) conductors through a reliable and easily scalable technique such as the PIT process combined with the melt process developed earlier by our group. Only few works on the fabrication of the 11 superconducting wires have been published. The best results in terms of transport properties were obtained in FeSe wires[18] realized through the so-called *in-situ* Fe-diffusion process, in which the Fe sheath works as precursor source. The resulting wires showed a reacted layer on the inside wall of the Fe sheath and a large hole at the centre of the core, where TeSe (or only Se) precursors were inserted before the heat treatment. Exploring the application of the *ex-situ* method is justified by the need of having a conductor with a good homogeneity, high fill-factor and useful mechanical properties. Moreover, we focused our work on the Te-doped 11 phase with the aim to get conductors with higher $T_C$ and better performances.

Our first investigation has concerned the metallic sheath, which in this process does not act as precursor as in the *in-situ* route, but just as mechanical support for the superconducting phase. Due to the strong reactivity of Se/Te with most metals (as evidenced by the large number of intermediate phases formed by Se and Te with any metal[4-12]) the Fe(Se,Te) phase reacted with all the tested sheaths. Our experiment led to establish that iron, although ferromagnetic and with non-optimal thermal conductivity, is the sole suitable metal in terms of chemical compatibility and mechanical properties. Iron is a good choice because it does not add extra elements to the superconducting phase. However, as reported in table 2, no matter what was the iron amount in the starting powders, an iron excess in the superconducting core was revealed.

Moreover, independently of the applied heat treatment and of the starting powder composition, the wire filament exhibits the coexistence of two isostructural phases with different Se/Te ratio, the dark one (close to $Fe_1Se_{0.9}Te_{0.1}$) and the bright one ($Fe_{1+y}$(Se,Te) with Se/Te ≈ 1). This is perfectly in line with what found in ref. 38 on analogous polycrystalline samples sintered at 700°C; moreover, a similar phase separation was observed also in melted $FeSe_xTe_{1-x}$ samples for x>0.5 and was ascribed to a miscibility gap in the Fe(Se,Te).[39] Recently the investigation of the full phase diagram (0≤x≤1) have been performed on thin films grown by pulsed laser ablation (PLD).[42] It has been found that thin films do not show phase separation in the range 0.6 ≤ x ≤ 0.8 and, in the newly discovered domain, improved superconducting properties have been observed. This result is even more significant because the targets with Se concentrations in the range of 0.6 ≤ x ≤ 0.8 present the phases A (like our bright phase) and B (like our dark phase). In the biphasic region, the ratio between the A and B components decreases linearly with increasing of the Se content.[43] This is in perfect agreement with our results. In fact we found that by changing the powder starting composition the ratio between the bright and dark phase varies: an increase in Se led to a corresponding increasing in the percentage of the dark minority phase.

Thus, we conclude that phase separation occurring in the range 0.6≤x≤0.8 is common to all the bulk samples, independently of the kind of synthesis and thermal treatment; this phenomenon can be avoided only by adopting thermodynamic non-equilibrium conditions, such as growth in thin films deposition. In our case the phase separation is combined with the excess of Fe provided by the sheath in the bright phase, and also this effect is independent of the starting composition and of the heat treatment.

Looking at the resistive transitions of Fig. 6, it can be noted that increasing the dark phase percentage, the superconducting properties are improved too. Indeed it was observed that wires with a dark phase percentage of about 10÷15% showed only partial superconducting transition, while wires with a complete transition have a dark-phase percentage around 30% or more, corresponding to a starting Se content higher than 30%. This observation, together with the fact that the $T_C$ for all samples is around 8-9 K, leads us to conclude that the superconducting behaviour of these wires is due to the dark Se-rich phase, whose composition is very close to the superconducting phase $FeSe_{1-\delta}$ with zero resistance at T = 8K.[40,22] Therefore, we can conclude that an amount of 30% of dark phase, which is close to the percolation limit, is necessary to get a complete superconducting transition. On the other hand, the $FeSe_{0.5}Te_{0.5}$ phase, which being the phase with the best superconducting properties was the object of our research, turns into the bright phase ($Fe_{1.02(2)}Se_{0.56(3)}Te_{0.44(3)}$) and the excess of iron coming from the sheath definitely suppresses its superconductivity.

The critical current density measurements on the melted wires show that the higher is the Se content, the higher is the $I_C$ value. Normalizing to the whole core (except for the sheath), the maximum measured value of $J_C$ is about 100 A/cm$^2$, definitely far from applicative level, but nevertheless the very same of that reported by Mizuguchi *et al.*[17] and about half of the best one for Fe(Se,Te) measured by Ozaki *et al.*[15] Both $J_C$ were calculated just over the cross sectional area of the reacted layer except for the holes.

In the light of what have discussed so far and assuming the dark Se-rich phase as the unique superconducting phase, we can actually recalculate our $J_C$ over the dark phase fraction area more appropriately. The so calculated values (labelled $J_{c\_D}$) are reported in table 3: the best value is now 400 A/cm$^2$, about 4 times higher than the $J_{c\_TOT}$ value.

It is interesting to compare our results with those reported by Ozaki *et al.* on a nominal $FeSe_{0.5}Te_{0.5}$ wire[15] which exhibited a reacted layer consisting of a slightly Se-substituted FeTe layer and a slightly Te-substituted FeSe layer, the first almost non-superconducting. The authors concluded that the transport $J_C$ could be improved by increasing the FeSe layer. This hypothesis was confirmed in a second experiment carried out by the same group[18] on a FeSe wire realized through the same method. In this case the reaction layer was made only by FeSe phase and a fourfold increase in $J_C$ is obtained.

Differently from the Ozaki works, we tried to move progressively from $FeSe_{0.5}Te_{0.5}$ to FeSe by changing the powder composition. However, the formation of FeSe inside the Fe(Se,Te) phase prevented the formation of a phase with a fixed Se/Te ratio. Hence, we reached the same conclusion that inside the iron sheath only the FeSe phase preserves the superconducting properties.

**V. CONCLUSIONS**

This work is focused on the investigation of the feasibility condition of PIT processed wires of Fe(SeTe) superconductor and their characterization in terms of microstructure and superconducting properties. In order to combine high in-field performances together with an industrial appealing fabrication process, the *ex-situ* method was proposed.

Our main results are the following: i) the only metal sheath compatible with the superconducting phase is iron but its contraindication is to provide an infinite iron source for the inner superconducting core; ii) after the heat treatment the starting powders de-mix in a bright phase ($Fe_{1+y}(Se,Te)$ with Se/Te ≈ 1), not superconducting because of the iron excess, and in a dark phase (close to $FeSe_{1-y}$), superconducting with $T_c$~8 K; iii) only the dark phase is superconducting and responsible for super-current transport. Since the FeSe phase has poorer superconducting properties ($T_C$, $J_C$, $H_{C2}$) than Te substituted compounds, the evaluated $J_c$ is quite low. However, our results are completely in line with those obtained by PIT method, thus we provide confirmation of difficulties in developing the PIT approach and we conclude that the outstanding $J_c$, $H_{c2}$ reported for the $FeSe_{0.5}Te_{0.5}$ cannot be easily reproduced in a wire.

Further investigation can be attempted to avoid or at least limit the excess of iron. Indeed, this is a crucial point that have been addressed in single crystals and polycrystalline samples with several strategies going from annealing, to electrochemical reaction techniques and so on. Several of them have been successful, but unfortunately they cannot be easily applied to the superconducting filament embedded within the sheath. In this contest, it could be interesting to test the effects of the addition to the starting powder of some elements, such as Carbon or Oxygen, which easily react with Fe reducing its content in the phase.

On the other side, to avoid the use of Fe we should develop more sophisticated metal sheaths combined with complex heat treatment sequences. However, as the PIT method is complicated by the use of complex procedures and expensive sheath, the interest on its use drops, and we might as well turn to the more sophisticated but far more effective conductors, such as the coated conductors. Example of $FeSe_{0.5}Te_{0.5}$-coated conductors have been already reported. [43] Thin films growth by PLD with a $CeO_2$ buffer, on coated conductor substrates exhibit critical current densities >$10^6$ A cm$^{-2}$ and exceeding $10^5$ A cm$^{-2}$ under 30 tesla magnetic fields at 4.2 K, which are much higher than those of low-temperature superconductors. Moreover, as reported in ref. 42 the films grown by PLD are not affected by phase separation as the bulks, and the film composition may be optimized in order to increase the critical temperature further.

Finally, we conclude that the problematic results of our broad study, together with the promising results obtained on Fe(Se,Te) films, push towards the development of iron-chalcogenide-coated conductors, which could be very attractive for high-field applications at liquid helium temperatures.

**ACKNOWLEDGMENTS**

This work has been supported by FP7 European project SUPER-IRON (grant agreement No.283204) and by Compagnia di Sanpaolo.

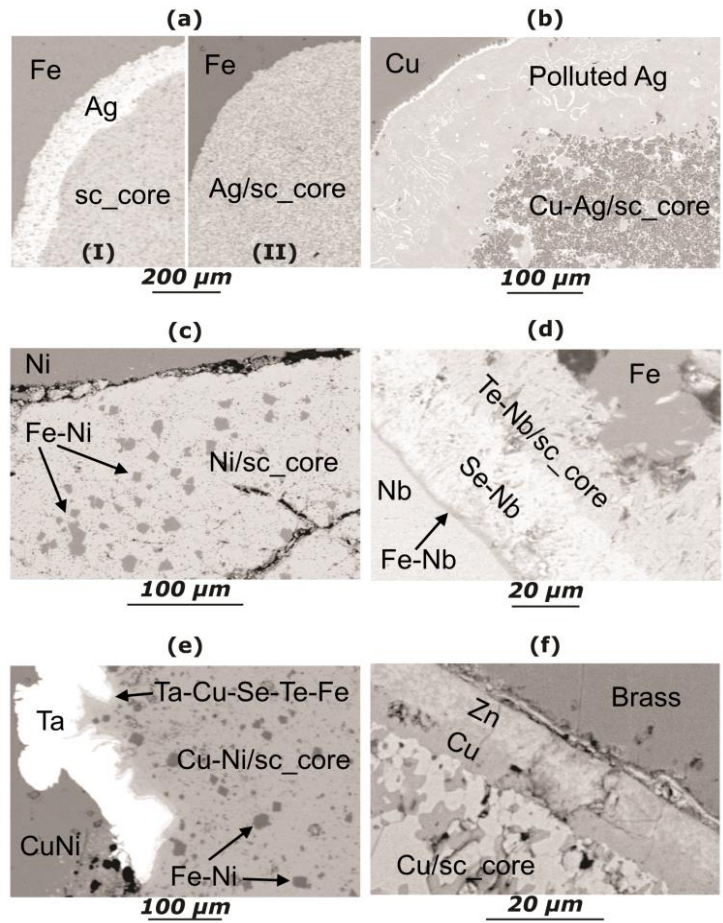

FIG. 1. Microstructures of wires with different sheaths. From top to bottom: (a) Fe-sheathed wire with Ag barrier before (I) and after (II) the heat treatment at 800°C; (b) Cu-sheathed wire with Ag barrier, heat-treated at 850°C; (c) Ni-sheathed tape, heat-treated at 850°C; (d) Nb-sheathed wire, heat-treated at 850°C; (e) Cupronickel 90-10-sheathed wire with broken tantalum barrier, heat-treated at 700°C; (f) Brass-sheathed wire, heat-treated at 850°C.

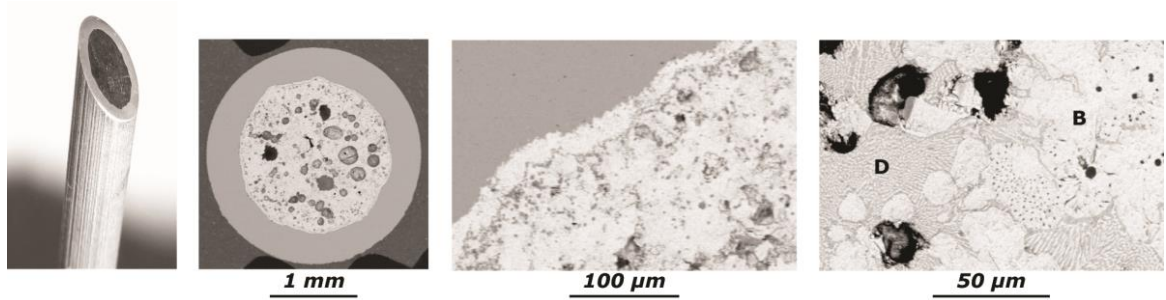

FIG. 2. Perspective view of the Fe-sheathed wire filled with FeSe$_{0.5}$Te$_{0.5}$ and corresponding SEM-BSE images of the micrographic cross section; the labels "D" and "B" indicate the dark and bright phases respectively.

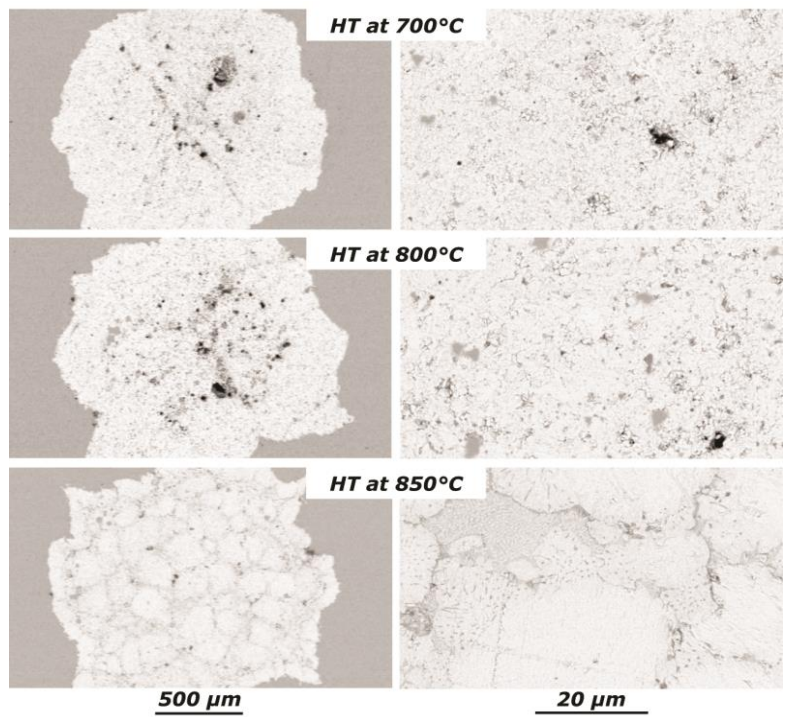

FIG. 3. SEM-BSE for Fe-sheathed wires filled with FeSe$_{0.5}$Te$_{0.5}$ powder, which underwent HT at different temperatures.

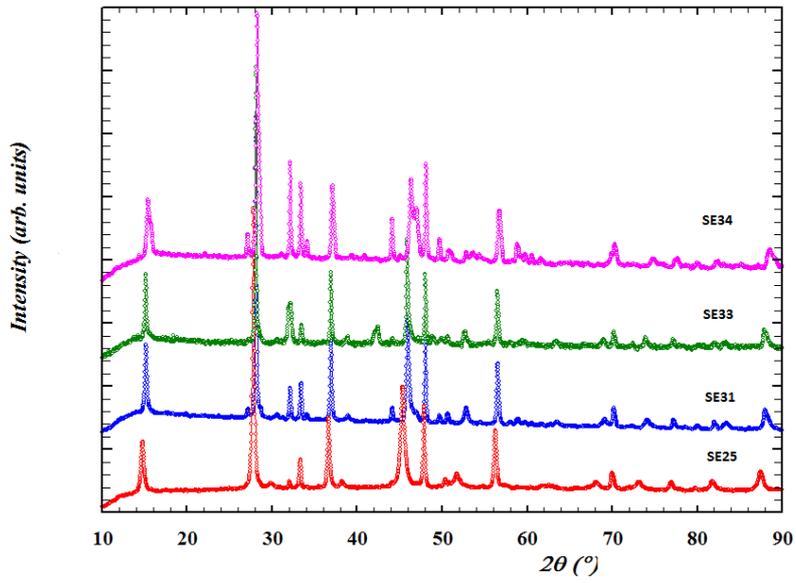

FIG. 4. XRD spectra for the as-prepared powders listed in Table 1. As a reference, the pattern calculated for FeSe$_{0.5}$Te$_{0.5}$ (PbO-type), is reported. The main peaks corresponding to the extra phases are also highlighted

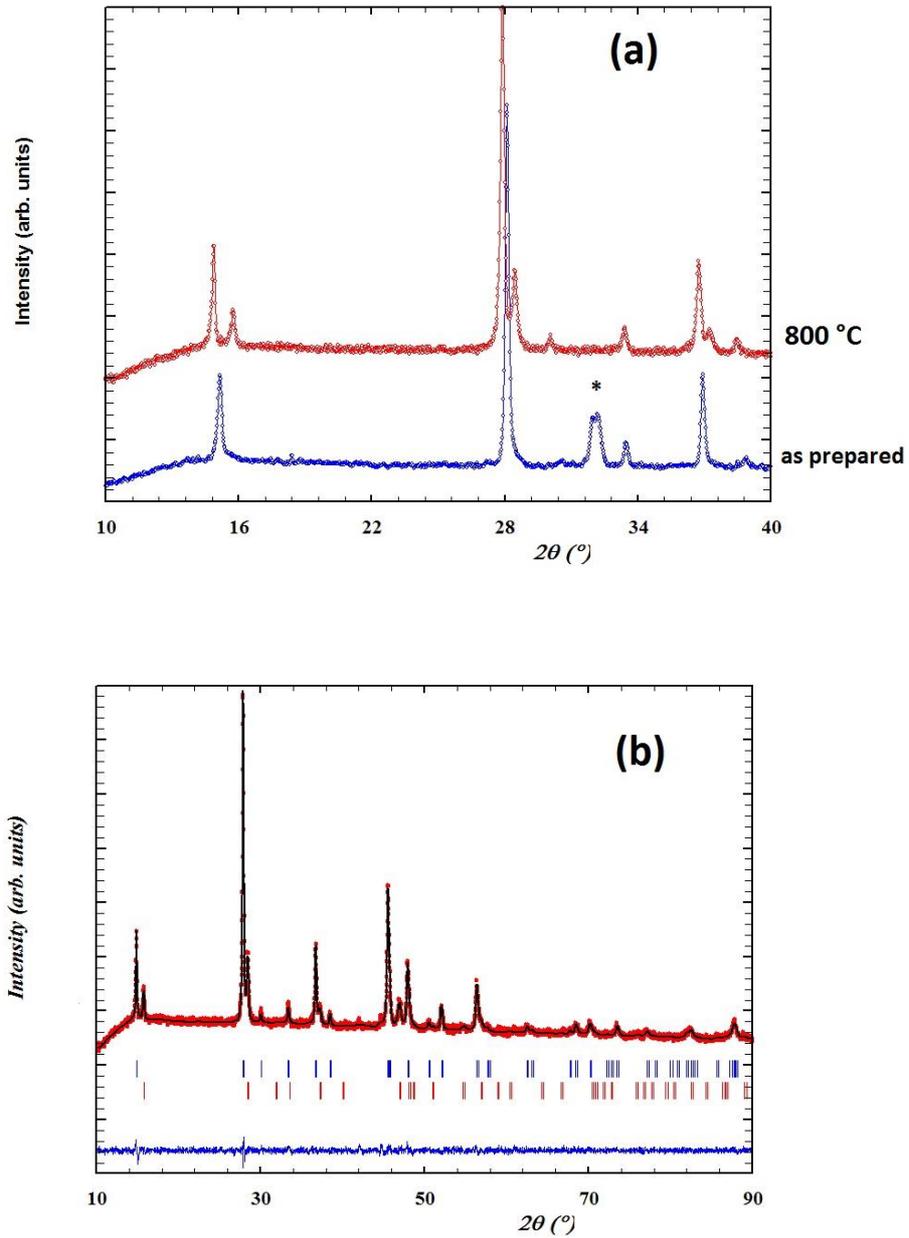

FIG. 5. XRD spectra of sample SE33. (a) As-prepared powders in comparison with wire-extracted powders; impurity peaks are marked by an asterisk. (b) Rietveld refinement plot of SE33 after heat treatment at 800°C; vertical bars correspond to the calculated Bragg peaks of $FeSe_{0.59}Te_{0.41}$ (main phase, $R_B$ = 2.86, upper row) and of $FeSe_{0.95}Te_{0.05}$ ($R_B$ = 2.62, lower row). The line at the bottom is the difference curve between observed and calculated data (Rp = 1.14, Rwp = 1.56, $\chi^2$ = 1.74).

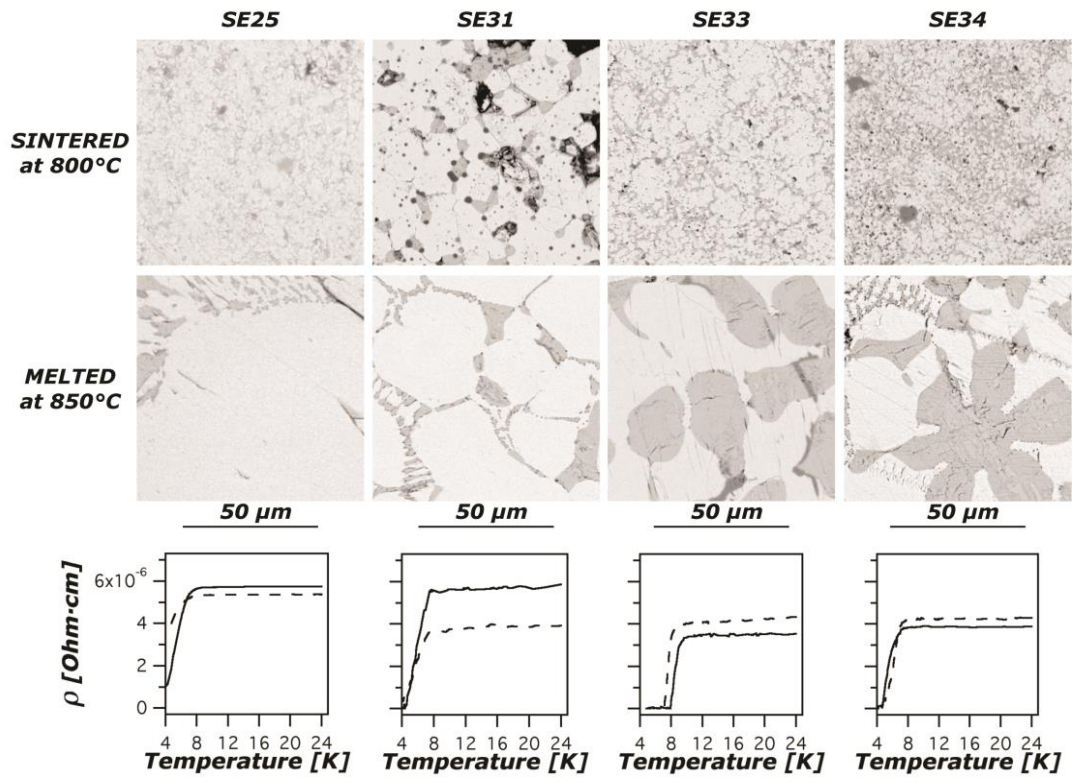

FIG. 6: Resistivity vs. temperature curves and corresponding SEM-BSE images at the same magnification (800x) for 800°C-sintered (dashed lines) and melted (solid lines) Fe-sheathed wires related to the powder contained in each wire.

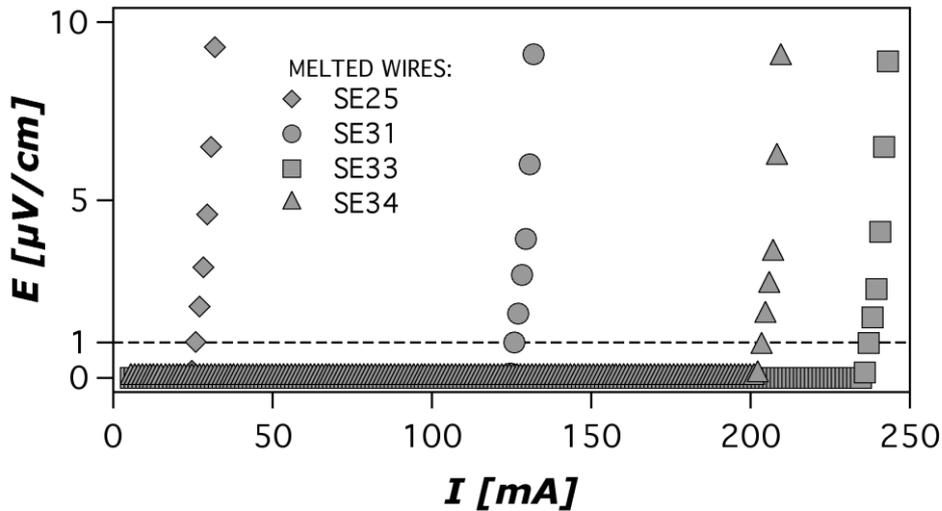

FIG. 7. V-I curves measured at 4.2 K for Fe-sheathed wires melted at 850°C and containing different powders.